\documentstyle[12pt,psfig,epsf]{article}
\newcommand{\be}{\begin{equation}}
\newcommand{\ee}{\end{equation}}

\newcommand{\beqa}{\begin{eqnarray}}
\newcommand{\eeqa}{\end{eqnarray}}

\newcommand{\eqref}[1]{(\ref{#1})}

\def\boxit#1{\vbox{\hrule\hbox{\vrule\kern8pt
\vbox{\hbox{\kern8pt}\hbox{\vbox{#1}}\hbox{\kern8pt}}
\kern8pt\vrule}\hrule}}
\def\mathboxit#1{\vbox{\hrule\hbox{\vrule\kern8pt\vbox{\kern8pt
\hbox{$\displaystyle #1$}\kern8pt}\kern8pt\vrule}\hrule}}

\def\IB{\relax\hbox{$\inbar\kern-.3em{\rm B}$}}
\def\IC{\relax\hbox{$\inbar\kern-.3em{\rm C}$}}
\def\ID{\relax\hbox{$\inbar\kern-.3em{\rm D}$}}
\def\IE{\relax\hbox{$\inbar\kern-.3em{\rm E}$}}
\def\IF{\relax\hbox{$\inbar\kern-.3em{\rm F}$}}
\def\IG{\relax\hbox{$\inbar\kern-.3em{\rm G}$}}
\def\IGa{\relax\hbox{${\rm I}\kern-.18em\Gamma$}}
\def\IH{\relax{\rm I\kern-.18em H}}
\def\IK{\relax{\rm I\kern-.18em K}}
\def\IL{\relax{\rm I\kern-.18em L}}
\def\IP{\relax{\rm I\kern-.18em P}}
\def\IR{\relax{\rm I\kern-.18em R}}
\def\IZ{\relax\ifmmode\mathchoice
{\hbox{\cmss Z\kern-.4em Z}}{\hbox{\cmss Z\kern-.4em Z}}
{\lower.9pt\hbox{\cmsss Z\kern-.4em Z}} {\lower1.2pt\hbox{\cmsss
Z\kern-.4em Z}}\else{\cmss Z\kern-.4em Z}\fi}

\def\II{\relax{\rm I\kern-.18em I}}

\begin{document}

\hfill  NRCPS-HE-01-26

\vspace{5cm}
\begin{center}
{\large \bf Weyl Invariant Gonihedric Strings
}

\vspace{2cm}

{\sl G.K.Savvidy\footnote{email:~savvidy@mail.demokritos.gr} and
R.Manvelyan\footnote{Permanent Address: Yerevan Physics Institute,
email:~manvel@moon.yerphi.am}}\\
National Research Center Demokritos,\\
Ag. Paraskevi, GR-15310 Athens, Hellenic Republic\\

\end{center}
\vspace{60pt}

\centerline{{\bf Abstract}}

\vspace{12pt}

\noindent We study quantum corrections in the earlier proposed
string theory, which is based on Weyl invariant purely extrinsic
curvature action. At one-loop level it remains Weyl invariant
irrespective of the dimension D of the embedding spacetime. To
some extent the counterterms are reminiscent of the ones in pure
quantum gravity. At classical level the string tension is equal to
zero and quarks viewed as open ends of the surface are propagating
freely without interaction. We demonstrate that quantum
fluctuations generate nonzero area term (string tension).


\newpage

\pagestyle{plain}
\section{Introduction}


In \cite{geo,sav} authors suggested so-called gonihedric model of
random surfaces, which is based on the concept of extrinsic
curvature. It differs in two essential points from the models
considered in the previous studies
\cite{polykov,kleinert,helfrich,weingarten}: first, it claims that
extrinsic curvature term alone should be considered as the
fundamental action of the theory
\be\label{funaction}
S =m \cdot L= m \int d^{2}\zeta
\sqrt{g}\sqrt{K^{ia}_{a}K^{ib}_{b}},
\ee
here $m$ has dimension of mass, $K^{i}_{ab}$ is a second
fundamental form (extrinsic curvature) and there is $no~area$ term
in the action. Secondly, it is required that the dependence on the
extrinsic curvature should be such that the action will have
dimension of length $ L ~~ \propto ~~length $, that is
proportional to the linear size of the surface similar to the path
integral \footnote{This is in contrast with the previous proposals
when the extrinsic curvature term is a dimensionless functional
$S(extrinsic~curvature) \propto 1$, invariant only under $rigid$
scale transformations.}. When the surface degenerates into a
single world line the functional integral over surfaces will
naturally transform into the Feynman path integral for point-like
relativistic particle
\be\label{limit}
S =m ~L~~ \rightarrow ~~ m \int ds .
\ee
At the classical level the string tension in this theory is equal
to zero and quarks viewed as open ends of the surface are
propagating freely without interaction $T_{classical}=0$ because
the action (\ref{funaction}) is equal to the perimeter of the flat
Wilson loop $ S \rightarrow  m(R+T)$. It was demonstrated in
\cite{geo} that quantum fluctuations generate the area term $A$ in
the effective action
\be\label{qtension}
S_{eff} = m~L~~ + ~~T_{q} ~ A~~ +~~ ...,
\ee
with the non-zero string tension  $T_{q}= \frac{D}{a^2}~(1 - ln
\frac{D}{\beta}) , $ here $D$ is the dimension of the spacetime,
$a$ is a scaling parameter. In the scaling limit $\beta
\rightarrow \beta_{c}=D/e$ the string tension has a finite limit.
Therefore at the tree level the theory describes free quarks with
string tension equal to zero, instead quantum fluctuations
generate nonzero string tension and, as a result, quark
confinement. The theory may consistently describe asymptotic
freedom and confinement as it is expected to be the case in QCD.

Dynamical string tension $T_q$  has been found in the discrete
formulation of the theory when the action (\ref{funaction}) is
written for the triangulated surfaces. Our aim now is to show that
this theory is well defined in one-loop approximation and that
similar generation of non-zero string tension takes place in a
continuum formulation of the theory and therefore may lead to a
nontrivial string theory. Here we shall treat quantum fluctuations
in two different ways following the works of Polyakov
\cite{polykov} and Kleinert \cite{kleinert} \footnote{See also
subsequent publications
\cite{pawelczyk,polchinski,kleinert1,diamantini,parthasarathy,plyushchay}
}.

\section{Weyl Invariance}
We shall represent the gonihedric action (\ref{funaction}) in a
Weyl invariant form
\begin{equation}\label{conaction}
S= m\int d^{2}\zeta \sqrt{g}\sqrt{ \left(\Delta(g)
X_{\mu}\right)^{2}},
\end{equation}
here ~$g_{ab}=\partial_{a}X_{\mu}\partial_{b}X_{\mu}$ ~is induced
metric,~$\Delta(g)= 1/\sqrt{g}~\partial_{a}\sqrt{g}g^{ab}
\partial_{b} $ ~is a Laplace operator and
$K^{ia}_{a}K^{ib}_{b}=\left(\Delta(g) X_{\mu}\right)^{2}$. The
second fundamental form $K$ is defined through the relations:
\begin{eqnarray}
K^{i}_{ab}n_{\mu}^{i}&=&\partial_{a}\partial_{b}X_{\mu}-
\Gamma^{c}_{ab}\partial_{c}X_{\mu}=\nabla_{a}\partial_{b}X_{\mu},\quad
\label{extrinc}\\
\quad n_{\mu}^{i}n_{\mu}^{j}&=&\delta_{ij},\quad n_{\mu}^{i}
\partial_{a}X_{\mu}=0,\label{normals}
\end{eqnarray}
where $n_{\mu}^{i}$ are $D-2$ normals and $a,b=1,2; \qquad
\mu=0,1,2,...,D-1; \quad i,j=1,2,...,D-2.$

We can introduce  independent metric coordinates $g_{ab}$ using
standard Lagrange multipliers $\lambda^{ab}$ and then fix
conformal gauge $g_{ab}=\rho\eta_{ab}$ using reparametrization
invariance. After fixing conformal gauge we shall have
\begin{equation}\label{conforgauge}
S=m \int d^{2}\zeta \left\{\sqrt{\left(\partial^{2}
X^{\mu}\right)^{2}}+\lambda^{ab}\left(\partial_{a}X^{\mu}\partial_{b}X^{\mu}-
\rho\eta_{ab}\right)\right\}.
\end{equation}
As one can see the first term is $\rho$ independent and therefore
is explicitly Weyl invariant ($local$ scale transformations). Note
that extrinsic curvature part of the Polyakov-Kleinert action
$S_{PK}={1\over e}\int d^2 \zeta \sqrt{g}K^{ia}_{a}K^{ib}_{b} =
{1\over e}\int d^2 \zeta \rho^{-1}(\partial^{2} X^{\mu})^{2}$ is
invariant only under $rigid$ scale transformations. This action
leads to the following equation of motion:
\be\label{acsel}
\partial^{2}\frac{\partial^{2}X_{\mu}}
{\sqrt{\left(\partial^{2}X_{\nu}\right)^2}}
-2~\partial_{a}\left(\lambda^{ab}\partial_{b}X_{\mu}\right)=0,~~~
\partial_{a}X_{\mu}\partial_{b}X_{\mu}-\rho\eta_{ab}=0,~~~
\lambda^{aa}=0.
\ee
In the light cone coordinates
$\zeta^{\pm}=(\zeta^{0}\pm\zeta^{1})/\sqrt{2}$ the conformal gauge
looks like:
\be\label{confgauge}
g_{++}=g_{--}=0, \qquad g_{+-}=\rho,
\ee
the connection has only two nonzero components $
\Gamma^{+}_{++}=\partial_{+}\ln{\rho},~
\Gamma^{-}_{--}=\partial_{-}\ln{\rho}$ and \\metric variation is:
\be
\delta g_{\pm\pm}=\nabla_{\pm}\varepsilon_{\pm}=
\rho\partial_{\pm}\varepsilon^{\mp}\label{nabla}.
\ee
In these coordinates the partition function takes the following
form:
\begin{equation}\label{partfunc}
  Z=\int \exp{\imath \{S\}}\det{\{\nabla_{+}\}}
  \det{\{\nabla_{-}\}}DX^{\mu}D\lambda_{ab}D\rho ,
\end{equation}
where $\det{\{\nabla_{+}\}}\det{\{\nabla_{-}\}}$  are
Faddeev-Popov determinants corresponding to the conformal gauge
(\ref{confgauge}).

To obtain quantum correction to the classical action we have to
expand our action around some classical solution
$(\bar{X}^{\mu},\bar{\lambda}^{ab},\bar{\rho})$~ up to second
order on small fluctuations
$(X^{\mu}_{1},\lambda^{ab}_{1},\rho_{1})$, thus $S= \bar{S}+
S_{2}+S_{int}$,~ where~$\bar{S}= m\int d^{2}\zeta
\sqrt{\bar{n}^{2}}$,
\beqa\label{peraction}
S_{2}=\int {m \over 2\sqrt{\bar{n}^{2}}}~ X^{\mu}_{1}
\partial^{4} \left(\eta^{\mu\nu}
-\frac{\bar{n}_{\mu}\bar{n}_{\nu}}{\bar{n}^{2}}
\right)X^{\nu}_{1}d^{2}\zeta + m \int (
2\lambda^{ab}_{1}~\bar{e}^{a}_{\mu}~\partial_{b}X^{\mu}_{1}-
\lambda^{aa}_{1}\rho_{1} ) d^{2}\zeta\nonumber,
\eeqa
\be\label{inter}
S_{int}=m\int d^{2}\xi
(\bar{\lambda}^{ab}\partial_{a}X^{\mu}_{1}\partial_{b}X^{\mu}_{1}+
\lambda^{ab}_{1}\partial_{a}X^{\mu}_{1}\partial_{b}X^{\mu}_{1}
+...),
\ee
where we have introduced convenient notations
$$\bar{n}_{\mu}=\partial^{2} \bar{X}_{\mu}
,~~~~~\bar{e}^{a}_{\mu}=\partial_{a}\bar{X}_{\mu}.$$ Here we admit
slow behavior of the first and second derivatives of the classical
solution $\bar{X}^{\mu}$ and natural separation of interaction
with external field $\bar{\lambda}^{ab}$.

One can see from (\ref{peraction}) that in one-loop approximation
we have factorization of tangential modes of $X^{\mu}_{1}$ and
cancellation of their contribution with the corresponding ghost
determinant. To see that, we shall expand $X^{\mu}_{1}$ into
tangential and normal fields $\phi_{a},~\xi^{i}$:
\begin{eqnarray}\label{expansion}
X_{1\mu}&=&\phi_{a}~\bar{e}^{a}_{\mu} + \xi^{i} ~\bar{n}_{\mu}^{i},\\
\bar{n}^{i}_{\mu}~\bar{e}^{a}_{\mu}&=&0, \quad
\bar{e}^{a}_{\mu}~\bar{n}_{\mu}=0.
\end{eqnarray}
The last relation, together with the following ones $
\bar{n}_{\mu}\bar{n}^{i}_{\mu}=\bar{K}^{ia}_{a}\equiv
\bar{K}^{i},~\bar{n}_{\mu}\bar{n}_{\mu}
=\bar{K}^{ia}_{a}~\bar{K}^{ib}_{b} \equiv \bar{K}^2, $ can be
easily  seen in conformal gauge (\ref{confgauge}). We also have
\be\label{action2}
\bar{S}= m \int d^{2}\zeta \sqrt{\bar{K}^{2}}.
\ee
We have to substitute expansion (\ref{expansion}) into
(\ref{peraction}) and take into account slow behavior of classical
fields compared with quantum fields:
\begin{eqnarray}\label{action3}
S_{2}= \int d^{2}\zeta \left\{ { m\over 2\sqrt{\bar{K}^{2}} }~
\xi^{i}~\partial^{4} \left( \delta^{ij}
-\frac{\bar{K}^{i}\bar{K}^{j}}{\bar{K}^{2}}
\right) \xi^{j}\right.\nonumber\\
+\lambda^{\pm\pm}_{1}\nabla_{\pm}\phi_{\pm}&+&\left.\lambda^{+-}_{1}
\rho_{0}\left(\partial_{+}
\phi^{+}+\partial_{-}\phi^{-}\right)-\lambda^{+-}_{1}
\rho_{1}\right\},\nonumber\\
S_{int}=m\int d^{2}\zeta
\bar{\lambda}^{ab}\left(\partial_{a}\xi^{i}\partial_{b}\xi^{i}+
\partial_{a}\phi^{c}\partial_{b}\phi_{c}\right).\label{interaction}
\end{eqnarray}
Using the last expression for $S_2$ in partition function
(\ref{partfunc}) and integrating it over $\rho_{1}$ and
$\lambda^{\pm\pm}_{1}$ we shall get delta functions
$\delta(\lambda^{+-}_{1}) \delta(\nabla_{\pm}\phi_{\pm})$. Then
integrating over $\lambda^{+-}_{1}$ and longitudinal components
$\phi^{\pm}$ we shall get determinants $det^{-1}{\nabla_{\pm}}$.
We observe now that in the one-loop approximation there is a
cancellation of these determinants with the ghost determinants in
(\ref{partfunc}) and therefore absence of conformal anomalies.
This should be verified in the next order where we have to
consider third order interactions of the form
$\lambda_{1}^{ab}\partial_{a}X^{\mu}_{1}\partial_{a}X^{\mu}_{1}$.

Finally we have the following one-loop partition function
\begin{equation}\label{finalaction}
Z_1=exp(~i ~\bar{S}~)~\int exp {\left\{~
i~S_{2}(\bar{K},\xi^{i})+i~S_{int}(\bar{\lambda},\xi^{i})
~\right\}}D\xi^{i},
\end{equation}
where
\begin{eqnarray}
S_{2}&=&\int d^{2}\zeta~\frac{m}{2
\sqrt{\bar{K}^{2}}}~\xi^{i}~\partial^{4}\left(\delta^{ij}
-\frac{\bar{K}^{i}\bar{K}^{j}}{\bar{K}^{2}} \right)
\xi^{j},\label{oneloop}\\
S_{int}&=&m\int d^{2}\zeta~
\bar{\lambda}^{ab}~\partial_{a}\xi^{i}\partial_{b}\xi^{i}.\label{oneloopint}
\end{eqnarray}
From quadratic part (\ref{oneloop}) we can deduce the propagator
\be\label{15}
\langle \xi^{i}(p)\xi^{j}(-p)\rangle
=\frac{\sqrt{\bar{K}^{2}}}{m}{\Pi^{ij}\over (p^2)^2},~~~~~~~
\Pi^{ij}=\delta^{ij}-\frac{\bar{K}^{i}\bar{K}^{j}}{\bar{K}^{2}},\quad
\Pi^{ii}=D-3
\ee
and calculate first correction to the classical action
(\ref{action2}). For that we have to contract $\xi^{i}$ fields in
(\ref{oneloopint}) using (\ref{15})
\begin{equation}\label{17}
W_{1}=m\int d^{2}\zeta~
\bar{\lambda}^{ab}~\langle\partial_{a}\xi^{i}\partial_{b}\xi^{i}\rangle
=\frac{D-3}{2\pi}\log\left(\Lambda/\tilde{\Lambda}\right)\int
d^{2}\zeta ~\bar{\lambda}^{aa}\sqrt{\bar{K}^{2}},
\end{equation}
thus $m_1 = m ~+~\bar{\lambda}^{aa}~
\frac{D-3}{2\pi}~\log\left(\Lambda/\tilde{\Lambda}\right) $.  On
the classical trajectory (\ref{acsel}) we have
$\bar{\lambda}^{aa}=0$ and therefore there is no quantum
corrections on-mass shell S-matrix elements and the theory is
one-loop ultraviolet finite. The absence of quantum corrections at
one-loop level is very similar to the pure quantum gravity in four
dimensions \cite{'tHooft:bx}. One can expect that summation of all
one-loop diagrams in external field $\bar{\lambda}^{ab}$ may
generate nontrivial solution and condensation of Lagrange
multiplier $\bar{\lambda}^{aa}\neq 0$ on quantum level due to the
delicate mixture of ultraviolet and infrared divergences
\cite{coleman,savvidy,batalin,polyakov1}.

\section{One-loop Effective Action}
Our aim here is to sum up all one-loop diagrams in the
$\bar{\lambda}^{ab}$ background. For that we have to keep all
one-loop diagrams which are induced by the vertex
$\bar{\lambda}^{ab}~\partial_{a}\xi^{i}\partial_{b}\xi^{i}$.
Integrating (\ref{finalaction}) together with the interaction term
one can get \cite{savvidy,batalin}
\be
W_{1}= {i\over 2} ~Trln\bar{H},
\ee
where
\be\label{ham}
\tilde{H}_{ij} = \Pi_{ij} ~\left\{~(\partial^{2})^{2} - 2
\sqrt{\bar{K}^{2}} ~\bar{\lambda}^{ab}\partial_{a}\partial_{b}
~\right\}.
\ee
We are looking for the solution in the form
\be\label{ansatz}
\bar{\lambda}^{ab} = \lambda \sqrt{g} g^{ab}= \lambda ~\eta^{ab},
\ee
where $\lambda$ is a constant field. Then
\be\label{oper}
\tilde{H}_{ij} = \Pi_{ij} ~\left\{~ (\partial^{2})^{2} - 2
\sqrt{\bar{K}^{2}} ~\lambda~\partial^{2}~ \right\}.
\ee
It is convenient to introduce the notation $P_{\alpha} = i
\partial_{\alpha}$ and we shall factor this operator into two pieces
$\tilde{H} = H*H_0$, where
\be\label{twopices}
H_{ij} = -\Pi_{ij} \left(P^{2} + 2  \lambda~ \sqrt{\bar{K}^{2}}
\right),~~~~~ H_{0ij} = -\Pi_{ij}  P^{2}.
\ee
The effective action takes the form
\be\label{effective}
L_{1}=-{i \over 2}\int {ds \over s} Tr U(s) ~ - ~{i \over 2}\int
{ds \over s} Tr U_{0}(s),
\ee
where trace $Tr$ is over Lorentz and world sheet coordinates $
\zeta$ and
\be\label{umatrix}
U(s) = exp{(-i H s)},~~~~~~~~U_0(s) = exp{(-i H_0 s)}.
\ee
Taking the trace over Lorentz indexes and using the matrix element
$
(\zeta^{'}(s)|\zeta^{''}(0)) = (\zeta^{'} |e^{iP^2 s} |\zeta^{''})
= {1 \over 4\pi s} exp{\left( -i {(\zeta^{''} - \zeta^{'})^2 \over
4s} \right) }
$
one can get
\be\label{effetive1}
L_{1} =-{i \over 2}\int {ds \over s} \left( {D-3 \over 4\pi s}
e^{2i\lambda s \sqrt{\bar{K}^{2}} } +{D-3 \over 4\pi s} \right)
\ee
and after rotation of the contour by $s \rightarrow -is$ and
substraction of vacuum contribution we shall get
\be\label{effec2}
L_{1} ={D-3 \over 8\pi}\int {ds \over s^2} \left( e^{- 2\lambda s
\sqrt{\bar{K}^{2}}} -1 \right) + C\lambda.
\ee
The counterterm is equal to $C \lambda$ and we can fix it by
normalization condition \cite{coleman,savvidy}:
\be\label{normal}
{\partial L_{1} \over \partial \lambda} |_{\lambda = m} =
\sqrt{\bar{K}^{2}},
\ee
after which we can get ultraviolet and infrared finite effective
action in the form
\be\label{integral}
L_{1} =  \tilde{\lambda} /2 + {D-3 \over 8\pi}\int_{0}^{\infty}
{ds \over s^2} \left( e^{- \tilde{\lambda} s } -1 +
\tilde{\lambda} s e^{- \tilde{m}s } \right),
\ee
where $\tilde{\lambda} = 2~\lambda~ \sqrt{\bar{K}^{2}}$ and
$\tilde{m} = 2~m~ \sqrt{\bar{K}^{2}}$. This expression can be
integrated out and we shall get
\be\label{finale}
L_{1}~ = ~\lambda~\sqrt{\bar{K}^{2}}~ +  ~ {D-3 \over 4\pi}
~\lambda~ \sqrt{\bar{K}^{2}}~[ ~ln( { \lambda \over m } ) -1 ~]
\ee
with its new minimum at the point $ <\lambda> =  m ~ exp{(-{4\pi
\over D-3})}$. The dynamical string tension is equal therefore to
$T_q = m^2 exp{(-{4\pi \over D-3})}$ because
\be\label{minimum}
S_{eff} = m^{2}e^{-{4\pi \over D-3}} \int d^{2}\zeta \sqrt{g}
g^{ab}
\partial_{a}X^{\mu}\partial_{b}X^{\mu},
\ee
as one can see from (\ref{conforgauge}).

\section{Consideration in the Physical Gauge }
To study quantum effects from a different perspective we shall
consider only normal, physical  perturbation of the world sheet
$X_{\mu}(\zeta)$ \cite{kleinert,blaschke}
\be\label{per}
\tilde{X}_{\mu} =  X_{\mu} + \xi^{i} n^{i}_{\mu},
\ee
then the first derivative is
$
\partial_{a}\tilde{X}_{\mu} =  \partial_{a}X_{\mu} + \partial_{a}\xi^{i}
n^{i}_{\mu} + \xi^{i}\partial_{a} n^{i}_{\mu}
$
and the metric is equal to
\beqa\label{metric}
\tilde{g}_{ab}  = g_{ab}  + \xi^{i}\partial_{a}X_{\mu}\partial_{b}
n^{i}_{\mu} + \xi^{i}\partial_{b}X_{\mu}\partial_{a} n^{i}_{\mu} +
\partial_{a}\xi^{i} \xi^{j}n^{i}_{\mu}\partial_{b}n^{j}_{\mu}\\+
\xi^{i} \partial_{b} \xi^{j}\partial_{a} n^{i}_{\mu} n^{j}_{\mu}+
\partial_{a}\xi^{i} \partial_{b}\xi^{i}+ \xi^{i} \xi^{j}\partial_{a} n^{i}_{\mu}
\partial_{b} n^{i}_{\mu},
\eeqa
then it follows that
\be\label{inversmetric}
\tilde{g}^{ab}  = g^{ab}  + 2 K^{iab} \xi^{i} - \nabla^{a}\xi^{i}
\nabla^{b}\xi^{i} + 3 K^{ia}_{c} K^{jcb}\xi^{i}\xi^{j}
\ee
and that
\be\label{deter}
\tilde{g}^{1/2}  = g^{1/2}[1  - \xi^{i} K^{ia}_{a} +{1\over 2}
R^{ij} \xi^{i} \xi^{j}  + {1\over 2}g^{ab} \nabla_{a}\xi^{i}
 \nabla_{b}\xi^{j}],
\ee
where
\be\label{definition}
K^{i}_{ab} = \partial_{a} \partial_{b}X_{\mu}~n^{i}_{\mu},~~~~~~~~
 R^{ij} = K^{ia}_{a}K^{jb}_{b} - K^{ia}_{b}K^{jb}_{a}.
\ee
For the second derivative we have
$
\partial_{a}\partial_{b}\tilde{X}_{\mu} =
\partial_{a}\partial_{b}X_{\mu} + \partial_{a}\partial_{b}\xi^{i}
n^{i}_{\mu} + \partial_{b}\xi^{i}\partial_{a} n^{i}_{\mu}+
\partial_{a}\xi^{i}\partial_{b} n^{i}_{\mu}+\xi^{i}\partial_{a}\partial_{b}
n^{i}_{\mu}. $ In order to compute the variation of the extrinsic
curvature we have to find the perturbation of the normals $\delta
n^{i}_{\mu} = \tilde{n}_{\mu}^{i} - n^{i}_{\mu} $,~ where ~$
\tilde{n}_{\mu}^{i}   \partial_{a}\tilde{X}_{\mu}
=0,~~\tilde{n}^{i}_{\mu} \tilde{n}^{j}_{\mu}=\delta^{ij}. $ From
this it follows that
\be\label{varia}
\partial_{a} \xi^{i} + \delta n^{i}_{\mu} \partial_{a} X_{\mu} +
\delta n^{i}_{\mu} (\partial_{a} n^{j}_{\mu} \xi^{j} +n^{j}_{\mu}
\partial_{a}\xi^{j})=0,~~~~n^{i}_{\mu} \delta n^{j}_{\mu} +
n^{j}_{\mu} \delta n^{i}_{\mu} + \delta n^{i}_{\mu} \delta
n^{j}_{\mu} =0
\ee
and we can always represent $\delta n^{i}_{\mu}$ in the form
$\delta n^{i}_{\mu} = A^{ia} \partial_{a}X_{\mu} +
B^{ij}n^{j}_{\mu} . $ The solution of (\ref{varia}) is
\be\label{normvaria}
\delta n^{i}_{\mu}  = -\partial^{a} X_{\mu}( \partial_{a} \xi^{i}
+ K^{jb}_{a}\xi^{j}\partial_{b} \xi^{i} ) - {1 \over 2}
n^{j}_{\mu}\partial^{a}\xi^{j}\partial_{a}\xi^{i}.
\ee
With this formulas we can get the variation of the extrinsic
curvature up to the second order
\be\label{extrivar}
\tilde{K}^{i}_{ab}  =\left(\partial_{a}\partial_{b}X_{\mu} +
\partial_{a}\partial_{b}\xi^{i} n^{i}_{\mu} +
\partial_{b}\xi^{i}\partial_{a} n^{i}_{\mu}+
\partial_{a}\xi^{i}\partial_{b} n^{i}_{\mu}+\xi^{i}\partial_{a}\partial_{b}
n^{i}_{\mu}\right) (n^{i}_{\mu} +\delta n^{i}_{\mu}),
\ee
substituting (\ref{normvaria}) one can get for the variation
$\delta K^{i}_{ab}= \tilde{K}^{i}_{ab} - K^{i}_{ab}$
\beqa\label{extrivar1}
\delta K^{i}_{ab}=  \nabla_{a} \nabla_{b}\xi^{i} - K^{ic}_{a}
K^{j}_{cb}\xi^{j} +\nabla_{b}K^{jc}_{a} ~\xi^{j}\nabla_{c}
\xi^{i}\\+ ( K^{j}_{bc}  \nabla_{a} \xi^{j} + K^{j}_{ac}
\nabla_{b} \xi^{j})\nabla^{c} \xi^{i} -{1\over
2}K^{j}_{ab}\nabla^{c}\xi^{j} \nabla_{c}\xi^{i}.
\eeqa
Using (\ref{inversmetric}) we can get the trace of the extrinsic
curvature
\beqa\label{traceextrivar}
\delta K^{ia}_{a}=  \nabla^{a} \nabla_{a}\xi^{i} + K^{ic}_{a}
K^{ja}_{c}\xi^{j} +\nabla^{a}K^{j}_{ab} ~\xi^{j}\nabla^{b} \xi^{i}
+ 2 K^{j}_{ab}  \nabla^{a} \xi^{j}\nabla^{b} \xi^{i} - K^{i}_{ab}
~\nabla^{a}\xi^{j}\nabla^{b} \xi^{j}\\
 + K^{j}_{ab} K^{ib}_{c}
K^{nca}\xi^{j}\xi^{n} - {1\over 2}K^{ja}_{a}\nabla^{c}\xi^{j}
\nabla_{c}\xi^{i} +2 K^{j}_{ab}\xi^{j}\nabla^{a}
\nabla^{b}\xi^{i}.
\eeqa
Using the last expression it is easy to compute the first and the
second variations of the action (\ref{funaction}):
\be\label{firstvar}
\delta_1 A_{Gonihedric} = m \int d^{2}\zeta \sqrt{g}~ \xi^{i}\{
\delta^{ij} \triangle   +  K^{ib}_{a} K^{ja}_{b} -
\delta^{ij}~K^{2} \} { K^{ja}_{a} \over \sqrt{K^{2} }},
\ee
\beqa\label{goniaction}
\delta_2 A_{Gonihedric} ={m \over 2}\int d^{2}\zeta
\sqrt{g}{1\over \sqrt{K^{2}}} \{\triangle \xi^{i}\triangle \xi^{i}
- {K^{i}K^{j}\over K^2}\triangle \xi^{i}\triangle \xi^{i} \\+
K^{i} K^{j}\nabla^{c} \xi^{i}\nabla_{c} \xi^{j}+ K^{i} K^{i}
\nabla^{c}\xi^{j}\nabla_{c}\xi^{j} -2
K^{ib}_{a} K^{ja}_{b} \nabla^{c}\xi^{i} \nabla_{c}\xi^{j}\\
-2K^{i} K^{i}_{bc} \nabla^{b}\xi^{j} \nabla^{c}\xi^{j} +{2 \over
K^2} K^{n} K^{nb}_{a} K^{ja}_{b} K^{i}\nabla^{e} \xi^{i}
\nabla_{e} \xi^{j}\},
\eeqa
where $K^i=  K^{ia}_{a}$~and $ K^2 = K^{ia}_{a} K^{ib}_{b}=K^i
K^i$. The equation of motion is
\be
\{ \delta^{ij} \triangle   - R^{ij}\} { K^{j} \over \sqrt{K^{2} }}
=0,
\ee
where $R^{ij}=K^i K^j - K^{ib}_{a} K^{ja}_{b}$. It is easy to see
that
\be\label{pairing}
<\nabla_{a}\xi^{i}\nabla_{b}\xi^{j}> =
\delta_{ab}~\Pi^{ij}~{\sqrt{K^{2}}\over m} ~{1\over
2\pi}~\log\left(\Lambda/\tilde{\Lambda}\right),
\ee
where $\Pi^{ij} = \delta^{ij} - {K^{i}K^{j} \over K^2}$. Using
(\ref{pairing}) one can find one-loop contribution
\beqa
{m \over 2 \sqrt{K^{2}}} ~K^{i}K^{j} ~2 ~\Pi^{ij}~ {\sqrt{K^{2}}
\over m}{1\over 2\pi}~\log\left(\Lambda/\tilde{\Lambda}\right) \\
+{m \over 2 \sqrt{K^{2}}}~ K^2~ 2~ (D-3) ~{\sqrt{K^{2}} \over m}
{1\over 2\pi}~\log\left(\Lambda/\tilde{\Lambda}\right)\\
+{m \over 2 \sqrt{K^{2}}} ~(-2)~ K^{ib}_{a} K^{ja}_{b}~ 2~
\Pi^{ij}~{\sqrt{K^{2}} \over m} {1\over 2\pi}~\log\left(
\Lambda/\tilde{\Lambda}\right)\\
+{m \over 2 \sqrt{K^{2}}} ~(-2)~ K^{2}~  (D-3)~ {\sqrt{K^{2}}\over
m} ~{1\over 2\pi}~\log\left(\Lambda/\tilde{\Lambda}\right)\\
+{m \over 2 \sqrt{K^{2}}} ~{2\over K^2}~K^{n} K^{nb}_{a}
K^{ja}_{b}K^{i}~2~ \Pi^{ij}~{\sqrt{K^{2}} \over m} {1\over
2\pi}~\log\left( \Lambda/\tilde{\Lambda}\right)
\eeqa
The first and the last terms are equal to zero because
$\Pi^{ij}K^j =0$, the second and the fourth terms cancel each
other and we see that {\it in this theory the counterterms do not
depend on the dimension of the embedding spacetime}. Only the
third therm is nonzero
\be
-2~ K^{ib}_{a} K^{ja}_{b}~ \Pi^{ij}~{1\over 2\pi}~\log\left(
\Lambda/\tilde{\Lambda}\right)=2(K^i K^j - R^{ij})~
\Pi^{ij}~{1\over 2\pi}~\log\left( \Lambda/\tilde{\Lambda}\right)
=-2R ~{1\over 2\pi}~\log\left( \Lambda/\tilde{\Lambda}\right),
\ee
here we have used equation of motion $R^{ij}K^j =0$. Thus it is
proportional to the Euler characteristic of the surface and can be
neglected. This is completely consistent with our previous result
that the theory is one-loop ultraviolet finite.

It is instructive to compare this result with the quantum
corrections in the theory with dimensionless action
\cite{polykov,kleinert}. The second variation of the
Polykov-Kleinert action is
\beqa\label{PKaction}
\delta A_{PK} ={1\over 2e_0} \int d^{2}u \sqrt{g}\{\triangle
\xi^{i}\triangle \xi^{i} + K^{ia}_{a} K^{jb}_{b}\nabla^{c}
\xi^{i}\nabla_{c} \xi^{j}+ {1\over 2}K^{ia}_{a} K^{ib}_{b}
\nabla^{c}\xi^{j} \nabla_{c}\xi^{j}\\ -2 K^{ib}_{a} K^{ja}_{b}
\nabla^{c}\xi^{i} \nabla_{c}\xi^{j} -2 K^{ia}_{a} K^{i}_{bc}
\nabla^{b}\xi^{j} \nabla^{c}\xi^{j} \},
\eeqa
and the one-loop correction renormalizes the coupling constant
\be
{1\over e} = {1\over e_0} -{D\over
4\pi}\log\left(\Lambda/\tilde{\Lambda}\right).
\ee

\section{Acknowledgement} One of the authors (R.M.) is indebted to
the National Research Center Demokritos for kind hospitality. The
work of G. Savvidy was supported in part by the by EEC Grant no.
HPRN-CT-1999-00161. The work of  R. Manvelyan was supported by the
Hellenic Ministry of National Economy Fellowship Nato Grant and
partially by Volkswagen Foundation of Germany.

\vfill
\end{document}